\documentclass[%
12pt,
 superscriptaddress,
 amsmath,amssymb,
 aps,UTF8,
]{revtex4-1}

\usepackage{graphicx}
\usepackage{dcolumn}
\usepackage{bm}
\usepackage{xcolor}
\usepackage{url}
\usepackage{float}
\usepackage{tabularx}
\usepackage{lipsum}
\usepackage{array}

\usepackage{booktabs}

\usepackage{subfigure}
\usepackage{multirow}

\begin{document}


\title{Transvese momentum dependent parton distributions of pion at leading twist}
\author{Wei Kou}
\affiliation{Institute of Modern Physics, Chinese Academy of Sciences, Lanzhou 730000, China}
\affiliation{University of Chinese Academy of Sciences, Beijing 100049, China}

\author{Chao Shi}
\email{cshi@nuaa.edu.cn}
\affiliation{Department of Nuclear Science and Technology, Nanjing University of Aeronautics and Astronautics, Nanjing 210016, China}

\author{Xurong Chen}
\email{xchen@impcas.ac.cn}
\affiliation{Institute of Modern Physics, Chinese Academy of Sciences, Lanzhou 730000, China}
\affiliation{University of Chinese Academy of Sciences, Beijing 100049, China}

\author{Wenbao Jia}
\affiliation{Department of Nuclear Science and Technology, Nanjing University of Aeronautics and Astronautics, Nanjing 210016, China}


\begin{abstract}
We calculate the leading twist pion unpolarized transverse momentum distribution $f_1(x,k_T^2)$ and the Boer-Mulders function $h_1^\perp(x,k_T^2)$, using leading Fock-state light front wave functions (LF-LFWFs) based on Dyson-Schwinger and Bethe-Salpeter equations. These DS-BSEs based LF-LFWFs provide dynamically generated s- and p-wave components, which are indispensable in producing chirally odd Boer-Mulders function that has one parton spin flipped. Employing a non-perturbative SU(3) gluon rescattering kernel to treat the gauge link of the Boer-Mulders function, we thus obtain both TMDs at hadronic scale and then evolve them to the scale of $\mu^2=4.0$ GeV$^2$.  We finally calculate the generalized Boer-Mulders shift and find it to be in agreement with the lattice prediction.

\end{abstract}

\maketitle


\section{Introduction}
\label{sec:intro}
Multidimensional imaging of hadrons has excited a lot
of interest for the last several decades. Transverse momentum dependent parton distributions functions (TMDs) contain important information of the three-dimensional internal structure of hadrons, especially the spin-orbit correlations of quarks within them \cite{Sivers:1989cc,Boer:1997nt,Barone:2001sp,Collins:2003fm,Angeles-Martinez:2015sea}. The TMDs of the pion and nucleon, which are spin-$0$ and spin-$1/2$ respectively, have thus received extensive studies from phenomenological models \cite{Pasquini:2005dk,Pasquini:2008ax,Bacchetta:2008af,Pasquini:2014ppa,Noguera:2015iia,Shi:2018zqd} and lattice QCD \cite{Engelhardt:2015xja,Ebert:2019okf,LatticeParton:2020uhz,Li:2021wvl}. Experimentally, they can be studied
with the Drell-Yan (DY), or semi-inclusive deep-inelastic scattering (SIDIS) process for nucleon \cite{Bacchetta:2006tn,Wang:2017zym,Bacchetta:2017gcc,Scimemi:2017etj,Vladimirov:2019bfa,Bury:2020vhj,Bacchetta:2022awv,Cerutti:2022lmb}.

The Boer-Mulders function is a T-odd distribution, which was initially considered to vanish due to the time-reversal invariance of QCD \cite{Collins:1992kk}, but later it became clear that they could appear dynamically by the initial or final states interaction \cite{Brodsky:2002cx,Brodsky:2002rv}. In other words, the T-odd distribution does not vanish in the case of non-trivial gauge link, which is required by the field theory of color gauge invariance \cite{Collins:2002kn,Ji:2002aa,Belitsky:2002sm}. In addition, the gauge link makes the T-odd distribution process dependent and selects the opposite sign depending on the process from SIDIS to DY.

Experimental measurements of the unpolarized pion-induced DY scattering \cite{Pasquini:2014ppa,Wang:2017zym} cross section and the azimuthal asymmetries are based on the unpolarized TMD and the Boer-Mulders function  of pion as inputs, which had both been measured \cite{NA10:1986fgk,NA10:1987sqk,Conway:1989fs}. The azimuthal asymmetry has been observed experimentally, and the pion Boer-Mulders function is important for explaining these observations. In addition to this, little is known about experiments with meson TMD, although this may change with the new COMPASS collaboration program for meson-induced DY scattering \cite{COMPASS:2010shj,COMPASS:2017jbv}.

Theoretical calculation on Boer-Mulders function has also received much attention. The pion Boer-Mulders function has
been predicted in the antiquark spectator model \cite{Lu:2004au,Meissner:2008ay}, the light front constituent quark model \cite{Pasquini:2014ppa,Wang:2017zym,Wang:2018naw,Lorce:2016ugb}, the MIT bag model \cite{Lu:2012hh}, the Nambu-Jona-Lasinio (NJL) model \cite{Noguera:2015iia,Ceccopieri:2018nop} and light front holographic approach \cite{Ahmady:2019yvo}. Except for the approach discussed by Ref. \cite{Ahmady:2019yvo}, the other previous models consider the interpretation of the Boer-Mulders function by gluon rescattering under perturbative cases. In Ref. \cite{Gamberg:2009uk}, the authors have made a notable attempt to go beyond this perturbative approximation within the antiquark spectator framework. Meanwhile, lattice calculation related to the pion Boer-Mulders function can also be found in the literature \cite{QCDSF:2007ifr,Engelhardt:2015xja}.

Here we take the light front QCD framework, where the TMDs are determined through overlap representations in terms of light front wave functions (LFWFs) \cite{Diehl:2000xz,Diehl:2003ny}. In \cite{Shi:2018zqd}, we employed the DS-BSEs based LF-LFWFs and calculate the unpolarized TMD of pion in the chiral limit. Here we will employ the LF-LFWFs of pion at physical mass from \cite{Shi:2021nvg}, and calculate the two leading-twist TMDs for pion. Our focus here is the Boer-Mulders function. According to the overlap representation at leading Fock-state, the Boer-Mulders function is proportional to p-wave ($q\bar{q}$ spin parallel) components of the LF-LFWFs \cite{Ahmady:2019yvo}. Hence it provides a sensitive probe to the p-wave components inside pion. In the the light-cone constituent quark model, all spin configurations are generated by the Melosh rotation. A  modeling scalar function is then introduced to count in the dynamical effect \cite{Schlumpf:1994bc,Pasquini:2014ppa}. In the light front holographic approach, as the conventional holographic pion LFWF only has spin-antiparallel contribution \cite{Brodsky:2006uqa,Bacchetta:2017vzh}, the authors add spin-parallel terms that are modulated by a dynamical spin parameter $B$, with $B$ phenomenologically determined from pion decay constant and form factors \cite{Ahmady:2016ufq,Ahmady:2019yvo}.  Here in the DS-BSEs approach, the s- and p-wave LFWFs components are simultaneously determined from their parent Bethe-Salpeter wave function which are dynamically solved, so the ratio between them are fixed \cite{Shi:2021nvg}. Given Boer-Mulders function's sensitivity to p-wave components, it is thus worth investigating the prediction from DS-BSEs based LF-LFWFs on pion Boer-Mulders function.

This paper is organized as follows: In Sec. \ref{sec:LFWFs} we recapitulate the LF-LFWFs of pion from its
Bethe-Salpeter wave functions. We then introduce the pion unpolarized TMD and Boer-Mulders function in Sec. \ref{sec:TMD and gauge link}. The re-scattering kernel of the SU(3) non-abelian gluon is employed and used to compute the gauge link in the Boer-Mulders function. The evolution of pion TMDs and comparison to lattice QCD regarding the generalized Boer-Mulders shift \cite{Engelhardt:2015xja} are represented in Sec. \ref{sec:evolution}. Finally we conclude in Sec. \ref{sec:conclusion}.

\section{pion light front wave functions}
\label{sec:LFWFs}
In light front QCD hadron states can be
expressed as the superposition of Fock-state components
classified by their orbital angular momentum projection $l_z$ \cite{Belitsky:2002sm}. For the pion with valence quark $f$ and valence anti-quark $\bar{h}$
the minimal (2-particle) Fock-state configuration is given
by \cite{Jia:2018ary,Li:2017mlw} 
\begin{equation}
	\begin{aligned}
		|M\rangle=& \sum_{\lambda_1, \lambda_2} \int \frac{d^2 \bold{k_T}}{(2 \pi)^3} \frac{d x}{2 \sqrt{x \bar{x}}} \frac{\delta_{i j}}{\sqrt{3}} \\
		& \times \Psi_{\lambda_1, \lambda_2}\left(x, \bold{k_T}\right) b_{f, \lambda_1, i}^{\dagger}\left(x, \bold{k_T}\right) d_{\bar{h}, \lambda_2, j}^{\dagger}\left(\bar{x}, \overline{\bold{k}}_\bold{T}\right)|0\rangle,
	\end{aligned}
\label{eq:2-particle}
\end{equation}
where $\bold{k_T}=(k^x,k^y)$ is the transverse momentum of the quark $f$, $\overline{\bold{k}}_\bold{T}=-\bold{k_T}$, $x=\frac{k^+}{P^+}$ is the light-cone momentum
fraction of the active quark, and $\bar{x}=(1-x)$. The quark helicity is labeled by $\lambda_i=(\uparrow, \downarrow)$ and $\delta_{i j} / \sqrt{3}$ is the color factor. The $b^\dagger$ and $d^\dagger$ are the creation
operators for a quark and antiquark, respectively. The $\Psi_{\lambda_1, \lambda_2}$ are the LF-LFWFs of the pion that encode the non-perturbative internal dynamical information. Meanwhile, constrained by parity properties, the four $\Psi_{\lambda_1, \lambda_2}$'s can be expressed with two independent scalar amplitudes \cite{Ji:2003yj}
\begin{equation}
	\begin{array}{ll}
		\Psi_{\uparrow, \downarrow}\left(x, \bold{k_T}\right)=\psi_0\left(x, \bold{k_T}^2\right), & \Psi_{\downarrow, \uparrow}\left(x, \bold{k_T}\right)=-\psi_0\left(x, \bold{k_T}^2\right) \\
		\Psi_{\uparrow, \uparrow}\left(x, \bold{k_T}\right)=k_T^{-} \psi_1\left(x, \bold{k_T}^2\right), & \Psi_{\downarrow, \downarrow}\left(x, \bold{k_T}\right)=k_T^{+} \psi_1\left(x, \bold{k_T}^2\right)
	\end{array}
\label{eq:Psi}
\end{equation}
where $k_T^{\pm}=k^1 \pm i k^2$. The subscript $i$ in $\psi_i$ refers to the absolute value of orbital angular momentum between quark and antiquark projected onto the longitudinal direction. Note that in light front constituent quark model and modified holographic model \cite{Pasquini:2014ppa,Ahmady:2019yvo}, the $\psi_0(x,\bold{k_T}^2)$ and $\psi_1(x,\bold{k_T}^2)$ assumed the same functional form, which does not necessarily hold in a general case \cite{Ji:2003yj,Shi:2020pqe}. In the DS-BSEs approach, they are obtained from the Bethe-Salpeter wave function via the light front projections \cite{Shi:2018zqd,Mezrag:2016hnp,Xu:2018eii}
\begin{equation}
	\begin{aligned}
		\psi_0\left(x, \bold{k_T}^2\right)=& \sqrt{3} i \int \frac{d k^{+} d k^{-}}{2 \pi} \\
		& \times \operatorname{Tr}_D\left[\gamma^{+} \gamma_5 \chi(k, P)\right] \delta\left(x P^{+}-k^{+}\right),
	\end{aligned}
\label{eq:psi l0}
\end{equation}
\begin{equation}
	\begin{aligned}
		\psi_1\left(x, \mathbf{k_T}^2\right)=&-\sqrt{3} i \int \frac{d k^{+} d k^{-}}{2 \pi} \frac{1}{\bold{k_T}^2} \\
		& \times \operatorname{Tr}_D\left[i \sigma_{+i} k_T^i \gamma_5 \chi(k, P)\right] \delta\left(x P^{+}-k^{+}\right),
	\end{aligned}
\label{eq:psi l1}
\end{equation}
where the trace is over Dirac indices. Here we take the LFWFs of pion at the mass of 130 MeV from our earlier calculation \cite{Shi:2021nvg}, which is based on a realistic interaction model under the rainbow-ladder truncation. We note that starting with exactly same setup within DS-BSEs, the $\rho$ and $J/\psi$ LF-LFWFs had been extracted and well reproduced the diffractive meson production cross section within color dipole model \cite{Shi:2021taf}. The generalized parton distribution and leading twist time reversal even TMDs were also studied for light and heavy vector mesons \cite{Shi:2022erw,Shi:2023oll}.

\section{Transverse momentum dependent parton distributions}
\label{sec:TMD and gauge link}
In this section, we unify the momentum and impact parameter-related symbols as follows,
\begin{equation}
	|\mathbf{k_T}|=|\mathbf{k}|=k_T,\ \ 	|\mathbf{q_T}|=|\mathbf{q}|=q_T,\ \ |\mathbf{b_T}|=|\mathbf{b}|=b_T.
\end{equation}

\subsection{TMDs with LFWFs}
\label{subsec:form}
 For the pion, there are two twist-2 TMDs: unpolarized quark TMD, $f_1(x,k_T)$, and the polarized quark TMD, $h_1^\perp(x,k_T)$, also known as the Boer-Mulders function \cite{Boer:1997nt,Boer:1999mm}. The pion TMDs are derived from the quark correlation function \cite{Mulders:1995dh,Boer:1997nt,Bacchetta:2006tn}
\begin{equation}
	\begin{aligned}
		&\Phi_{i j}^{[\Gamma]}(x, \mathbf{k}) \\
		&=\int \frac{d z^{-} d^{2} z_{T}}{2 \pi(2 \pi)^{2}} e^{i z \cdot k}\left\langle\pi\left|\bar{\Psi}_{j}(0) \Gamma \mathcal{L}^{\dagger}(\mathbf{0} \mid n) \mathcal{L}(\mathbf{z} \mid n) \Psi_{i}(z)\right| \pi\right\rangle_{z^{+}=0},
	\end{aligned}
\label{eq:quark correlation}
\end{equation}
where the gauge link content is described as \cite{Mulders:1995dh,Boer:1997nt,Bacchetta:2006tn}
\begin{equation}
	\begin{aligned}
		&\mathcal{L}_{A^{+}=0} \left(\mathbf{z}_{T} \mid n\right) \\
		&=\mathcal{P} \exp \left(-i g \int_{\mathbf{z}_{T}}^{\infty} \mathrm{d} \eta_{T} \cdot \mathbf{A}_{T}\left(\eta^{-}=n \cdot \infty, \mathbf{z}_{T}\right)\right),
	\end{aligned}
\label{eq:gauge link}
\end{equation}
which guarantees colour gauge invariance and $n=(0,+1(-1),0)$ are appropriate for defining TMDs in SIDIS (Drell-Yan) processes. In particular, this reverses the sign of all T-odd distribution functions entering the correlator \cite{Pasquini:2014ppa}. The unpolarized TMD and Boer-Mulders function are given by \cite{Pasquini:2014ppa}
\begin{equation}
	f_{1}\left(x, k_{T}\right)=\frac{1}{2} \operatorname{Tr}\left(\Phi^{\left[\gamma^{+}\right]}\right),
	\label{eq:utmds}
\end{equation}
and 
\begin{equation}
	h_{1}^{\perp}\left(x, k_{T}\right)=\frac{\epsilon^{i j} \mathbf{k}^{j} M_{\pi}}{2 k_{T}^{2}} \operatorname{Tr}\left(\Phi^{\left[i \sigma^{i+} \gamma_{5}\right]}\right),
	\label{eq:bmf}
\end{equation}
respectively.

Without consideration of the gauge link, the trace of the quark correlation is written as \cite{Ahmady:2019yvo}
\begin{equation}
	\begin{aligned}
		\operatorname{Tr}\left(\Phi^{[\Gamma]}\right)=& \sum_{h, \bar{h}, h^{\prime}} \frac{1}{16 \pi^{3} k^{+}} \Psi_{h^{\prime} \bar{h}}^{*}(x, \mathbf{k}) \Psi_{h \bar{h}}(x, \mathbf{k}) \\
		& \times \bar{u}_{h^{\prime}}\left(k^{+}, \mathbf{k}\right) \Gamma u_{h}\left(k^{+}, \mathbf{k}\right).
	\end{aligned}
\label{eq:trace of quark correlation}
\end{equation}
For unpolarized TMD $f_1(x,k_T)$, one has $\Gamma=\gamma^+$ and
the light front matrix element \cite{Lepage:1980fj} is
\begin{equation}
	\bar{u}_{h^{\prime}}\left(k^{+}, \mathbf{k}\right) \gamma^{+} u_{h}\left(k^{+}, \mathbf{k}\right)=2 k^{+} \delta_{h h^{\prime}}.
	\label{eq:gamma_plus}
\end{equation}
From Eq. (\ref{eq:utmds}) one has
\begin{equation}
	f_{1}\left(x, k_{T}\right)=\frac{1}{16 \pi^{3}} \sum_{h, \bar{h}}\left|\Psi_{h \bar{h}}(x, \mathbf{k})\right|^{2}.
	\label{eq:utmd_lfwf_version}
\end{equation} 
In a parton model, integrating over $\mathbf{k}$ in $f_{1}\left(x, k_{T}\right)$ gives the familiar collinear valence parton distribution $f_1(x)$ \cite{Wang:2017onm}.

On the other hand, for the Boer-Mulders function (\ref{eq:bmf}), the light front matrix element is
\begin{equation}
	\bar{u}_{h^{\prime}}\left(k^{+}, \mathbf{k}\right) i\epsilon^{i j} k^{j} \sigma^{i+} \gamma^{5} u_{h}\left(k^{+}, \mathbf{k}\right)=2i k^{+} h^{\prime} \delta_{-h^{\prime} h} k_{T} e^{-i h^{\prime} \theta_{k_{T}}}.
	\label{eq:gamma5}
\end{equation}
Substituting Eqs.~(\ref{eq:2-particle},\ref{eq:Psi},\ref{eq:gamma5}) into Eq. (\ref{eq:trace of quark correlation}) one gets a vanishing Boer-Mulders function because 
\begin{equation}
	\sum_{h, \bar{h}} \Psi_{-h \bar{h}}^{*}(x, \mathbf{k}) h k_{T} e^{i h \theta_{k_{T}}} \Psi_{h \bar{h}}(x, \mathbf{k})=0.
	\label{eq:vanished bmf}
\end{equation} 
In this case, the gauge link must be taken into consideration. Dynamically, T-odd PDFs emerge from the gauge link structure of the multi-parton quark and/or gluon correlation functions \cite{Boer:2003cm,Brodsky:2002cx,Collins:2002kn,Belitsky:2002sm} which describe initial/final-state interactions (ISI/FSI) of the active parton via soft gluon exchanges with the target remnant \cite{Gamberg:2009uk}. Many studies have been performed to model the T-odd PDFs in terms of the FSIs where soft gluon rescattering is approximated by perturbative one-gluon exchange in Abelian models \cite{Brodsky:2002cx,Ji:2002aa,Goldstein:2002vv,Boer:2002ju,Gamberg:2003ey,Gamberg:2003eg,Bacchetta:2003rz,Lu:2004hu,Gamberg:2007wm,Bacchetta:2008af}. In Ref. \cite{Gamberg:2009uk}, the authors go beyond this approximation by applying non-perturbative eikonal methods to calculate higher-order gluonic contributions
from the gauge link while also taking into account color, which is collectively referred to as gluon
rescattering. In \cite{Ahmady:2019yvo}, the authors assumed the physics to be encoded in a gluon rescattering kernel $G\left(x, \mathbf{k}-\mathbf{k}^{\prime}\right)$ such that
\begin{equation}
	\begin{aligned}
		\operatorname{Tr}\left(\Phi^{[\Gamma]}\right)=& \sum_{h, \bar{h}, h^{\prime}} \int \frac{\mathrm{d}^{2} \mathbf{k}^{\prime}}{16 \pi^{3} k^{\prime+}} G\left(x, \mathbf{k}-\mathbf{k}^{\prime}\right) \Psi_{h^{\prime} \bar{h}}^{*}\left(x, \mathbf{k}^{\prime}\right) \\
		& \times \Psi_{h \bar{h}}(x, \mathbf{k}) \bar{u}_{h^{\prime}}\left(k^{\prime+}, \mathbf{k}^{\prime}\right) \Gamma u_{h}\left(k^{+}, \mathbf{k}\right).
	\end{aligned}
\label{eq:trace include G}
\end{equation}
Inserting Eq. (\ref{eq:gamma5}) to (\ref{eq:trace include G}), Eq. (\ref{eq:bmf}) yields
\begin{equation}
	\begin{aligned}
	h_{1}^{\perp}\left(x, k_{T}^{2}\right)=&\frac{ M_{\pi}}{k_{T}^{2} } \int \frac{\mathrm{d}^{2} \mathbf{k}^{\prime}}{16 \pi^{3}} i G\left(x, \mathbf{k}-\mathbf{k}^{\prime}\right) \\
		& \times \sum_{h, \bar{h}} \Psi_{-h, \bar{h}}^{*}\left(x, \mathbf{k}^{\prime}\right) h k_{T} e^{i h \theta_{k_{T}}} \Psi_{h, \bar{h}}(x, \mathbf{k}).
	\end{aligned}
\label{eq:main bmf}
\end{equation}
Determining the $G\left(x, \mathbf{k}-\mathbf{k}^{\prime}\right)$ thus would allow us to extract the Boer-Mulders function.

\subsection{Lensing function and gluon rescattering kernel}
\label{subsec:SU3}
An exact nonperturbative computation gluon rescattering kernel is yet not available and, in practice, some approximation scheme is necessary. Here we take the idea of chromodynamic lensing function introduced in \cite{Burkardt:2003uw}. It had been applied to the study of phenomenology of the Sivers asymmetry \cite{Bacchetta:2011gx} and  discussions on the its validity and limitation in model studies had been given in \cite{Pasquini:2019evu}. The authors of Ref. \cite{Gamberg:2009uk} show that one could apply non-perturbative eikonal methods to calculate higher-order gluonic contributions from the gauge link  to obtain the QCD lensing function \cite{Burkardt:2007xm} from the eikonal amplitude for quark-antiquark scattering via the exchange of both direct and crossed ladder diagrams of non-Abelian soft gluons.  According to the scheme of Ref. \cite{Gamberg:2009uk}, the lensing function in momentum space $I(x,q_T)$ connects the first moment of the Boer-Mulders
function with the chiral-odd pion generalized parton distribution (GPD) \cite{Meissner:2008ay}:
\begin{equation}
	M_{\pi}^{2} h_{1}^{\perp(1)}(x)=\int \frac{\mathrm{d}^{2} \mathbf{q}}{2(2 \pi)^{2}} q_{T} I\left(x, q_{T}\right) \mathcal{H}_{1}^{\pi}\left(x,-\left(\frac{q_{T}}{\bar{x}}\right)^{2}\right),
	\label{eq:1st-gpd}
\end{equation}
where the chiral-odd pion GPD is given by \cite{Meissner:2008ay}
\begin{equation}
	\mathcal{H}_{1}^{\pi}\left(x,-\Delta_{T}^{2}\right)=\frac{\epsilon^{i j} \Delta^{i} M_{\pi}}{2 \Delta_{T}^{2}} \int \frac{\mathrm{d} z^{-}}{2 \pi} e^{i k^{+} z^{-}}\left\langle P^{+}, \boldsymbol{\Delta}\left|\bar{\Psi}(0) \sigma^{i+} \gamma_{5} \Psi(z)\right| P^{+}, \mathbf{0}\right\rangle_{z^{+}=0}
	\label{eq:gpd}
\end{equation}
with $\mathbf{\Delta}=-\mathbf{q} / \bar{x}$ and $\mathbf{q}=\mathbf{k}-\mathbf{k}^\prime$ denoting the momentum transfer. As noted in Ref. \cite{Gamberg:2009uk}, the factorization (\ref{eq:1st-gpd}) does not hold in general \cite{Meissner:2008ay,Meissner:2009ww}. However, in Ref. \cite{Ahmady:2019yvo} the author have shown that the factorization also holds in the overlap representation with modified holographic LFWFs. Here we find it holds for our DS-BSEs based LFWFs as well. The lensing function and the gluon rescattering kernel are thus connected as \cite{Ahmady:2019yvo}
\begin{equation}
	i G\left(x, q_{T}\right)=-\frac{2}{(2 \pi)^{2}} \frac{\bar{x} I\left(x, q_{T}\right)}{q_{T}}.
	\label{eq:I and iG}
\end{equation}

The lensing function is derived for final state
rescattering by soft $U(1)$, $SU(2)$ and $SU(3)$ gluons. In Ref. \cite{Gamberg:2009uk}, the authors derive the eikonal amplitude for quark-antiquark scattering via the exchange of generalized
infinite ladders of gluons of the Abelian and non-Abelian case. In all three cases, the lensing function is negative. In the impact parameter space, the eikonal amplitude yields the lensing function of the form \cite{Gamberg:2009uk},
\begin{equation}
	\mathcal{I}\left(x, b_{T}\right)=\frac{\bar{x}}{2 N_{c}} \frac{\chi^{\prime}}{4} C\left(\frac{\chi}{4}\right),
	\label{eq:lensing_bt}
\end{equation}
\begin{equation}
	\begin{aligned}
		C\left[\frac{\chi}{4}\right] \equiv & {\left[(\operatorname{Tr} \mathfrak{I}[f])^{\prime}\left(\frac{\chi}{4}\right)+\frac{1}{2} \operatorname{Tr}\left[(\mathfrak{I}[f])^{\prime}\left(\frac{\chi}{4}\right)(\mathfrak{R}[f])\left(\frac{\chi}{4}\right)\right]\right.} \\
		&\left.-\frac{1}{2} \operatorname{Tr}\left[(\mathfrak{I}[f])\left(\frac{\chi}{4}\right)(\mathfrak{R}[f])^{\prime}\left(\frac{\chi}{4}\right)\right]\right],
	\end{aligned}
	\label{eq:color function}
\end{equation} 
where $\chi^\prime$ denotes the first derivative with respect to $b_T/\bar{x}$, and $(\mathfrak{I}[f])^{\prime}$ and $(\mathfrak{R}[f])^{\prime}$ are the first derivatives of the real and imaginary parts of the color function $f$. The eikonal phase $\chi$ is defined as the Hankel transformation of the gauge-independent part of the gluon propagator $\tilde{\mathcal{D}}_{1}\left(-k_{T}^{2}\right)$ \cite{Gamberg:2009uk},
\begin{equation}
	\chi\left(\frac{b_T}{\bar{x}}\right)=\frac{g^{2}}{2 \pi} \int_{0}^{\infty} d k_{T} k_{T} J_{0}\left(\frac{b_T}{\bar{x}} k_{T}\right) {\mathcal{D}}_{1}\left(-k_{T}^{2}\right),
	\label{eq:eikonal phase}
\end{equation}
where $J_0$ is a Bessel function of the first kind. The coupling $g$ represents the strength of the quark (antiquark) - gluon interaction.

For $SU(3)$ gluons, the real and imaginary parts of the color function $f$ are derived in Ref. \cite{Gamberg:2009uk}. One can get the power like forms with the numerical parameters \cite{Gamberg:2009uk}
\begin{equation}
	\begin{aligned}
		&\Re\left[f_{\alpha \beta}^{S U(3)}\right](a)=\delta_{\alpha \beta}\left(-c_{2} a^{2}+c_{4} a^{4}-c_{6} a^{6}-c_{8} a^{8}+\cdots\right), \\
		&\Im\left[f_{\alpha \beta}^{S U(3)}\right](a)=\delta_{\alpha \beta}\left(c_{1} a-c_{3} a^{3}+c_{5} a^{5}-c_{7} a^{7}+\cdots\right),
	\end{aligned}
\label{eq:su3 f}
\end{equation}
where $a=\chi/4$ and the parameters' values are set as $c_1=5.333$, $c_2=6.222$, $c_3=3.951$, $c_4=1.934$, $c_5=0.680$, $c_6=0.198$, $c_7=0.047$, $c_8=0.00967$ \cite{Gamberg:2009uk}.

In order to compute the eikonal phase, we follow the previous work \cite{Gamberg:2009uk} and use the non-perturbative Dyson-Schwinger gluon propagator, which has been given in Refs. \cite{Fischer:2002hna,Fischer:2003rp,Alkofer:2008tt},
\begin{equation}
	\begin{aligned}
		\mathcal{D}_{1}\left(k_{T}^{2}, \Lambda_{\mathrm{QCD}}^{2}\right)=& \frac{1}{k_{T}^{2}}\left(\frac{\alpha_{s}\left(k_{T}^{2}\right)}{\alpha_{s}\left(\Lambda_{\mathrm{QCD}}^{2}\right)}\right)^{1+2 \delta} \\
		& \times\left(\frac{c\left(k_{T}^{2} / \Lambda^{2}\right)^{\kappa}+d\left(k_{T}^{2} / \Lambda^{2}\right)^{2 \kappa}}{1+c\left(k_{T}^{2} / \Lambda^{2}\right)^{\kappa}+d\left(k_{T}^{2} / \Lambda^{2}\right)^{2 \kappa}}\right)^{2},
	\end{aligned}
\label{eq:DSE propagator}
\end{equation}
where the running strong coupling $\alpha_{s}$ presented in \cite{Fischer:2002hna}
\begin{equation}
	\alpha_{s}\left(\mu^{2}\right)=\frac{\alpha_{s}(0)}{\ln \left[\mathrm{e}+a_{1}\left(\mu^{2} / \Lambda^{2}\right)^{a_{2}}+b_{1}\left(\mu^{2} / \Lambda^{2}\right)^{b_{2}}\right]},
	\label{eq:as}
\end{equation}
The values for the fitting parameters are $\Lambda=0.71$ GeV, $a_1=1.106$, $a_2=2.324$, $b_1=0.004$ and $b_2=3.169$. Note that in solving the pion Bethe-Salpeter wave functions, we used a different interaction model. In that model, the full gluon propagator, the strong coupling constant and the full quark-gluon vertex are grouped and can not be separated. So here we employ Eq.~(\ref{eq:DSE propagator})  instead, which also allows  comparison with results from \cite{Ahmady:2019yvo}. 

Finally, the momentum space lensing function is given by the inverse Fourier transformation  \cite{Ahmady:2019yvo}
\begin{equation}
	I\left(x, q_{T}\right) \frac{\mathbf{q}^{i}}{q_{T}}=-\frac{i}{\bar{x}^{3}} \int \mathrm{d}^{2} \mathbf{b} \exp \left(-i \frac{\mathbf{q} \cdot \mathbf{b}}{\bar{x}}\right) \mathcal{I}\left(x, b_{T}\right) \frac{\mathbf{b}^{i}}{b_{T}}.
	\label{eq:lensing_moment}
\end{equation}
From Eq. (\ref{eq:I and iG}) to (\ref{eq:lensing_moment}), one can calculate the whole lensing function in momentum space $I\left(x, q_{T}\right)$ and eventually the gluon rescattering kernel $iG(x,q_T)$.

\subsection{$f_1(x,k_T)$ and $h_1^\perp(x,k_T)$ at initial scale $\mu=\mu_0$}
\label{subsec:Q0-TMD}

\begin{figure}[htbp]
	\centering  
	\subfigure{
		\label{fig:h1}
		\includegraphics[width=0.47\textwidth]{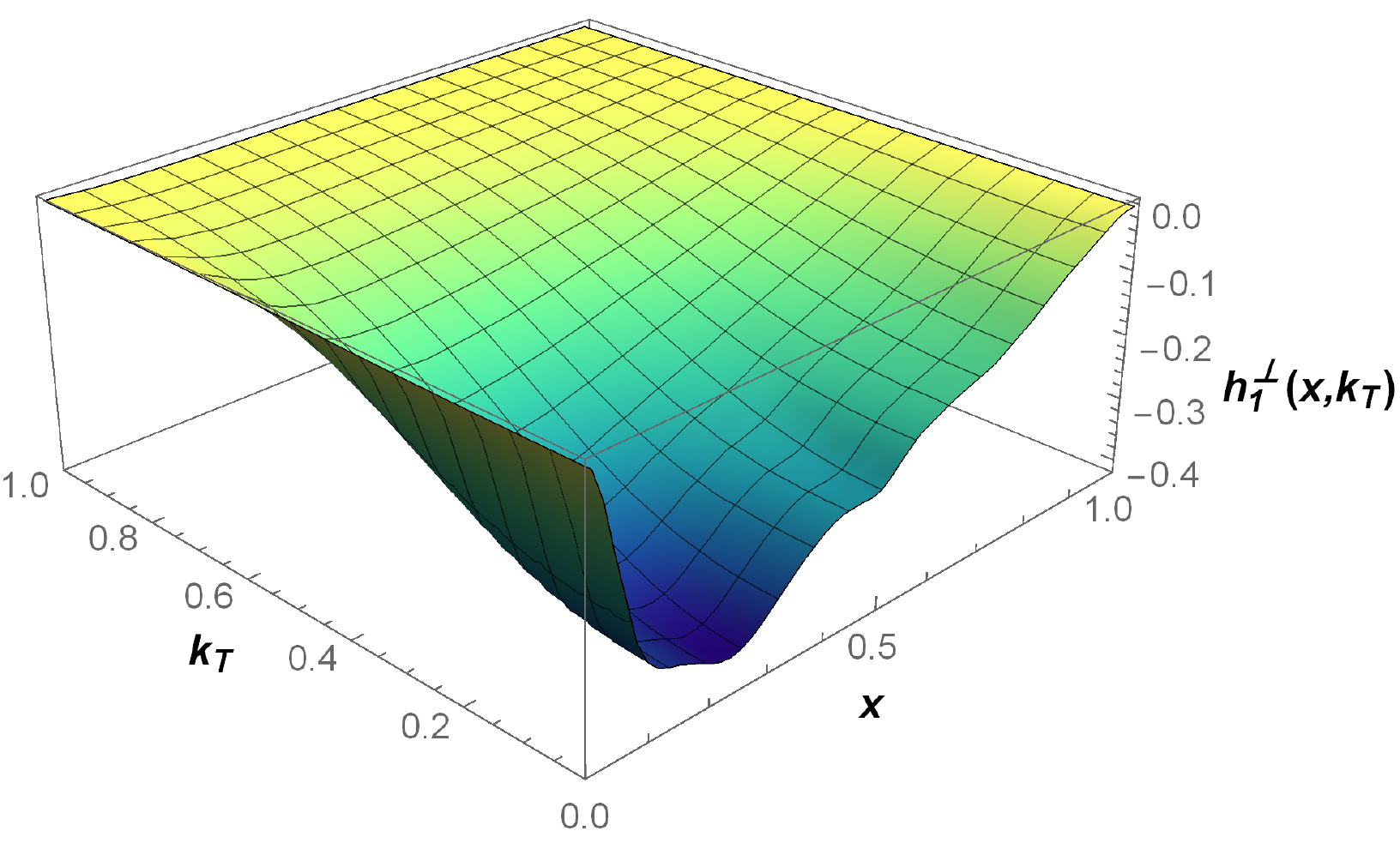}}
	\subfigure{
		\label{fig:f1}
		\includegraphics[width=0.47\textwidth]{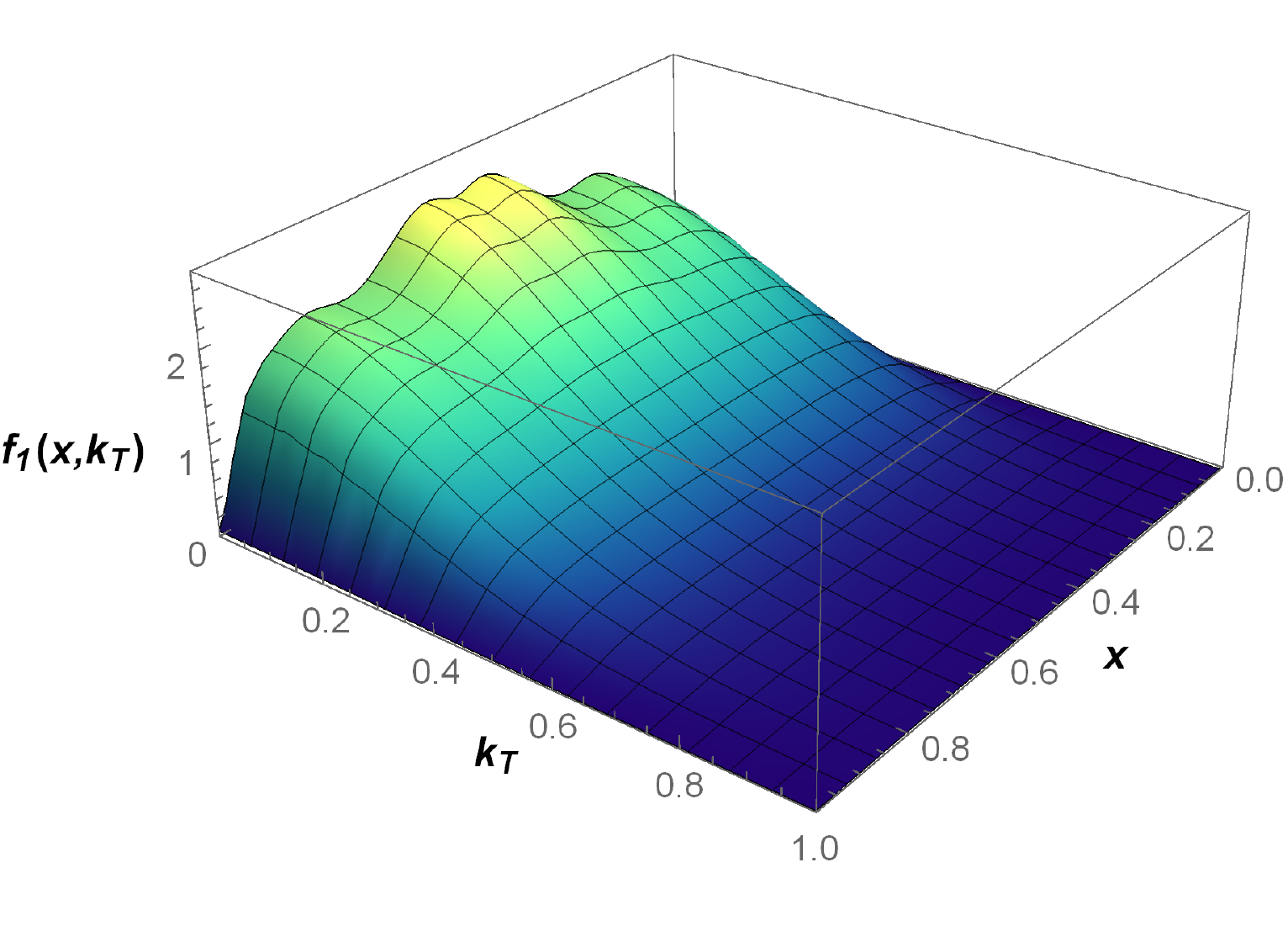}}
	\caption{(left) The Boer-Mulders function generated by the non-perturbative kernel. (right) The unpolarized TMD $f_1$.}
	\label{fig:TMDs}
\end{figure}

Having specified both the light front wave functions and
the gluon rescattering kernel, we are now in the position to study the pion TMDs. According to Eqs. (\ref{eq:utmd_lfwf_version}, \ref{eq:main bmf}), our calculated Boer-Mulders function and unpolarized TMD are shown in FIG. \ref{fig:TMDs}.

\begin{figure}[htbp]
	\centering
	\includegraphics[width=0.6\textwidth]{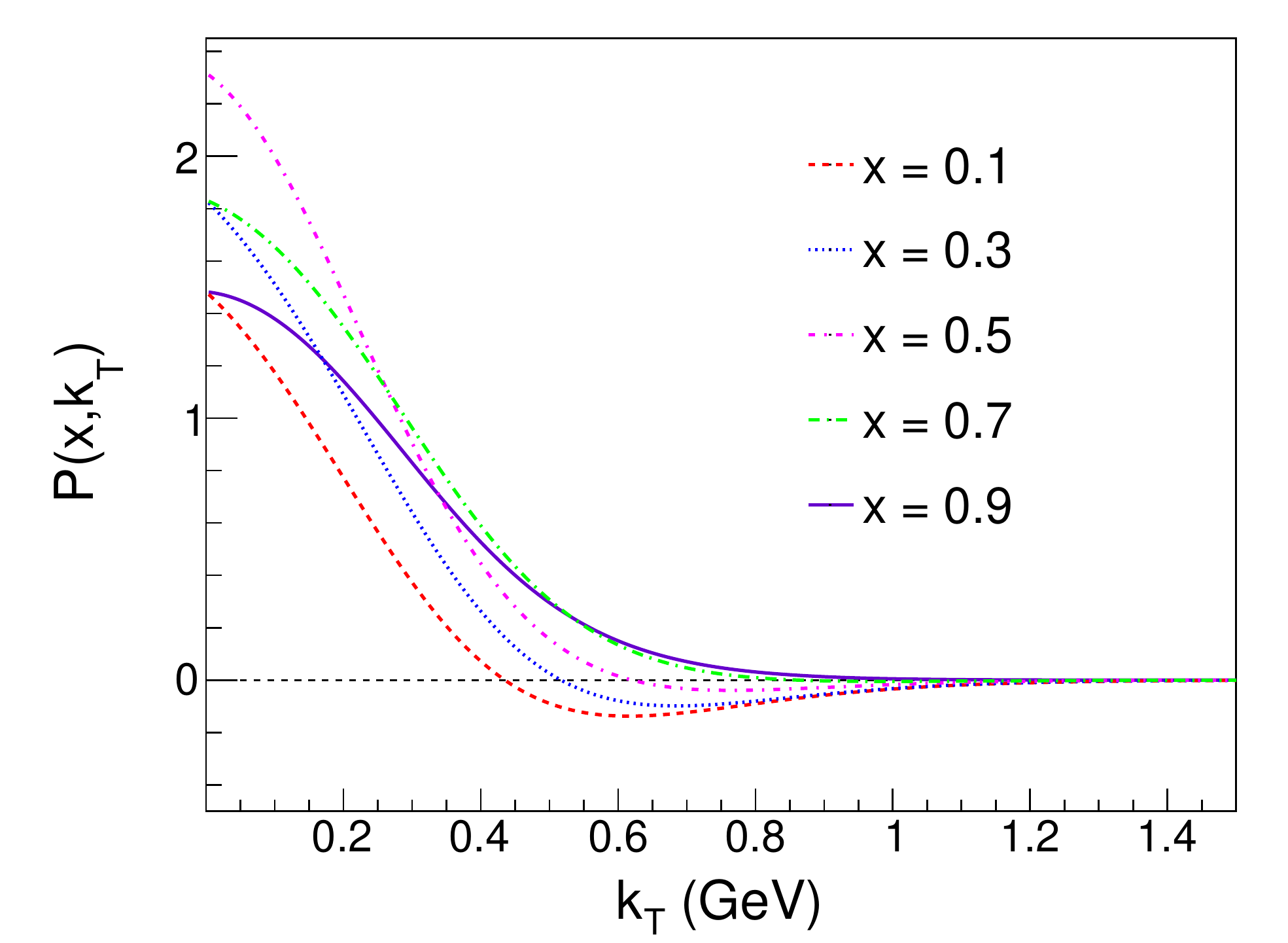}
	\caption{$P(x,k_T)$ with the Boer-Mulders function generated by the non-perturbative kernel at $x=0.1$, $x=0.3$, $x=0.5$, $x=0.7$ and $x=0.9$.
	}
	\label{fig:positive}
\end{figure}

We first check the general constraint of positivity bound condition \cite{Bacchetta:1999kz} using our TMDs
\begin{equation}
	P\left(x, k_{\perp}\right) \equiv f_{1}\left(x, k_{\perp}\right)-\frac{k_{\perp}}{M_{\pi}}\left|h_{1}^{\perp}\left(x, k_{\perp}\right)\right| \geq 0.
	\label{eq:positive}
\end{equation}
In FIG. \ref{fig:positive} we show that this constraint is mostly satisfied, with violations only occur for small $x$ at moderate and large $k_T$. We note that similar violations were reported in model studies \cite{Pasquini:2014ppa,Kotzinian:2008fe,Pasquini:2011tk,Wang:2017onm} with the perturbative kernel. They  indicate a limitation of current nonperturbative models to accurately capture the large $k_T$ behavior of the TMDs \cite{Ahmady:2019yvo}. Fortunately, the violation is generally small in magnitude, and insignificant for $x\gtrsim 0.5$. This is compatible with the leading Fock state truncation we employ, which works better in the valence region.

\section{The generalized Boer-Mulders shift}
\label{sec:evolution}
Parton distribution functions are defined within a certain
regularization scheme at a given renormalization scale. At some low ``hadronic scale" one deals with
constituent (valence) degrees of freedom carrying the total
hadron momentum: a constituent quark-antiquark pair in
the pion case, or three constituent quarks in the nucleon
case \cite{Pasquini:2014ppa}. The results from phenomenological studies refer to an assumed low initial scale $\mu_0$. The value of $\mu_0$ is not known a $\boldsymbol{priori}$, but can be determined by comparing the $\langle x\rangle-$Mellin moment of pion valence PDF with experimental extraction or lattice prediction. In this work, the initial scale is set to be $\mu_0=0.5$ GeV. In the following we shall assume that theoretical uncertainties
due to scheme dependence are smaller than the generic accuracy of LFWFs approaches. 

In order to compare with lattice calculation, we evolve our TMDs from $\mu^2=\mu_0^2=0.25$ GeV$^2$ to $\mu^2=4.0$ GeV$^2$. The later scale corresponds to the Lattice simulation in \cite{Engelhardt:2015xja}. In the parton model, the collinear unpolarized PDF is related to the unpolarized TMD as 
\begin{equation}
	f_1(x)=\int \mathrm{d}^2 \mathbf{k_T} f_1\left(x, k_{T}\right),
	\label{eq:collinear PDF}
\end{equation} 
which denotes the original PDF of pion and satisfies the Dokshitzer-Gribov-Lipatov-Altarelli-Parisi (DGLAP)
evolution \cite{Dokshitzer:1977sg,Gribov:1972ri,altarelli1977asymptotic}. The first $k_T$-moment of Boer-Mulders function in pion is defined as \cite{Wang:2017onm}
\begin{equation}
	h_{1 \pi}^{\perp(1)}(x)=\int d^2 \mathbf{k_T} \frac{\mathbf{k_T}^2}{2 M_\pi^2} h_{1 \pi}^{\perp}\left(x, {k_T}\right),
	\label{eq:BMF 1st moment}
\end{equation}
with the pion mass $M_\pi=130$ MeV. At the tree level, $h_{1 \pi}^{\perp(1)q}(x)$ can be related to the twist-3 quark-gluon correlation function $T_{q, F}^{(\sigma)}(x, x)$ \cite{Wang:2017onm}
\begin{equation}
	h_1^{\perp(1) q}(x) =\frac{1}{2 M_\pi} T_{q, F}^{(\sigma)}(x, x),
	\label{eq:twist-3}
\end{equation}
whereas the QCD evolution for $T_{q, F}^{(\sigma)}(x, x)$ is given in Ref. \cite{Zhou:2008mz,Kang:2012em}. 

\begin{figure}[htbp]
	\centering  
	\subfigure{
		\label{fig:xf1}
		\includegraphics[width=0.47\textwidth]{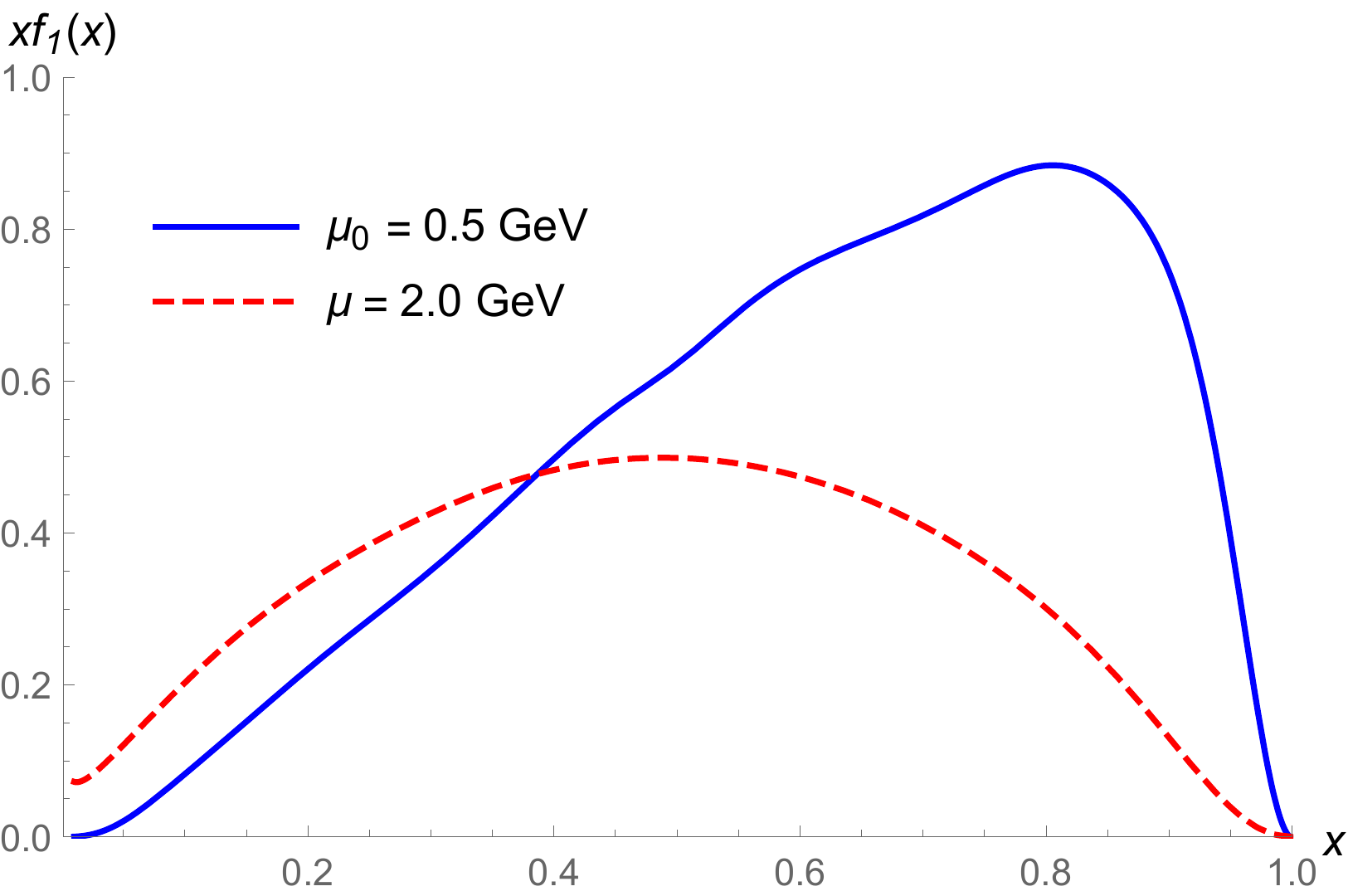}}
	\subfigure{
		\label{fig:xh1}
		\includegraphics[width=0.47\textwidth]{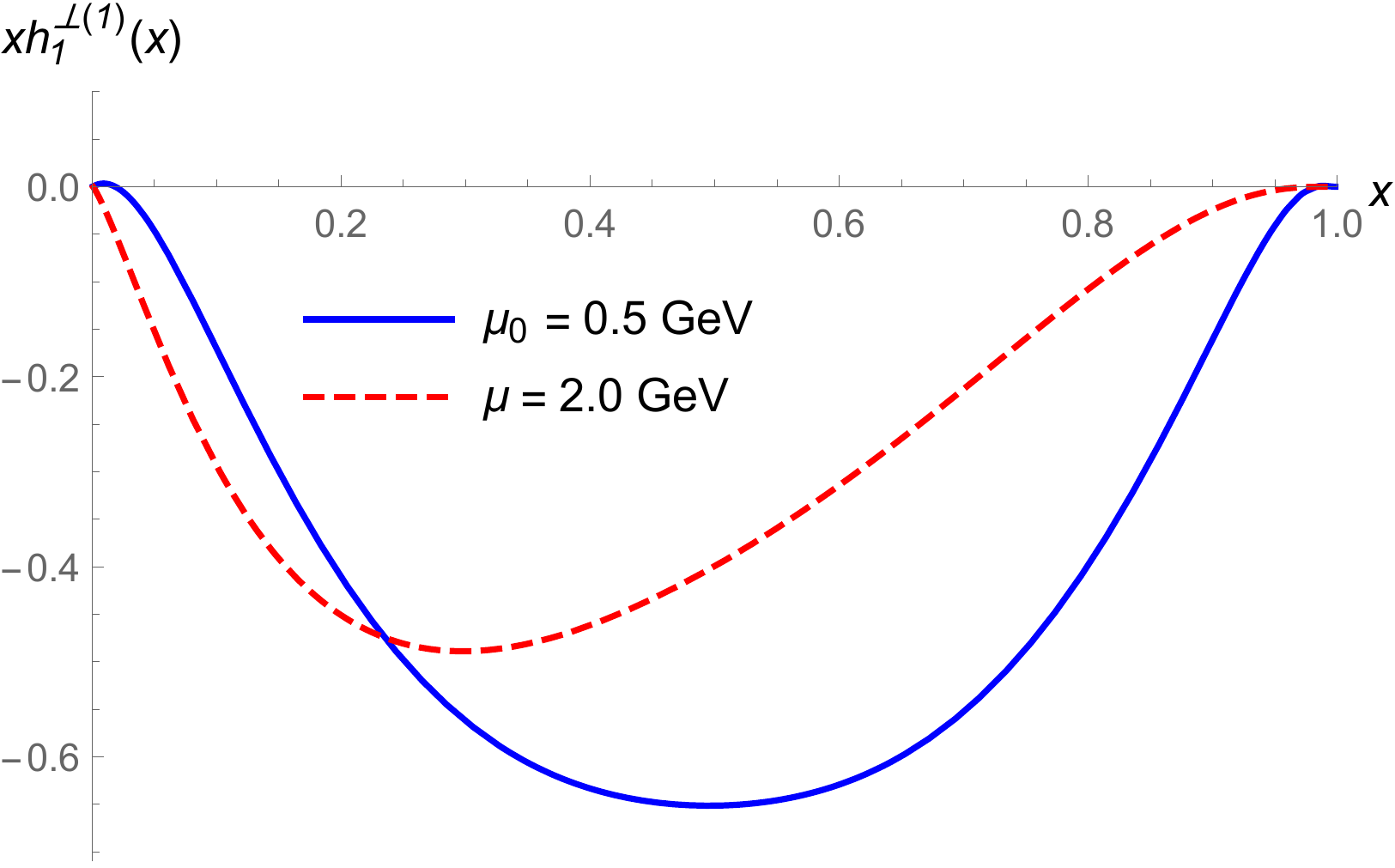}}
	\caption{(left) Comparison of $xf_1(x)$ at two different scales $\mu_0=0.5$ GeV (blue solid) and $\mu=2.0$ GeV (red dashed). (right) Comparison of $xh_{1 \pi}^{\perp(1)}(x)$ at two different scales $\mu_0=0.5$ GeV (blue solid) and $\mu=2.0$ GeV (red dashed).}
	\label{fig:evo-x}
\end{figure}

To perform the DGLAP evolution on PDF $xf_1(x)$, we adopt the QCDNUM \cite{Botje:2010ay} package at leading order, and we choose the strong coupling constant as $\alpha_{s}(M_Z^2)=0.118$. For the $h_1^{\perp(1) q}(x)$ case, we follow Ref. \cite{Wang:2017onm} and take the evolution kernel to be \cite{Wang:2017onm}
\begin{equation}
	P_{q q}^{h_1^{\perp(1)}}(x) \approx \Delta_T P_{q q}(x)-N_C \delta(1-x),
	\label{eq:BMF-kernel}
\end{equation}
with $\Delta_T P_{q q}(x)=C_F\left[\frac{2 z}{(1-z)_{+}}+\frac{3}{2} \delta(1-x)\right]$. To describe the evolution of $xh_{1 \pi}^{\perp(1)}(x)$, we also use the QCDNUM \cite{Botje:2010ay} program but insert the kernel (\ref{eq:BMF-kernel}) into the code. The numerical results are shown in FIG. \ref{fig:evo-x}. One can find that the peak of the evolved distribution shifts towards smaller $x$.

On the other hand, the TMDs evolution is more complicated \cite{Collins:2011zzd,Aybat:2011zv}. Taking the unpolarized TMD as an example, it is implemented in the space Fourier-conjugate
to $k_T$
\begin{equation}
	\widetilde{f}_1\left(x, b_{T}^2 ; \mu\right)=\int_0^{\infty} d k_{T} k_{T} J_0\left(b_{T} k_{T}\right) f_1\left(x, k_{T}^2 ; \mu\right).
	\label{eq:fourier}
\end{equation}
The TMDs generally depend on two scales, the ultraviolet renormalization scale $\mu$ and the $\zeta$ that regulates the rapidity divergence. Here,  we follow the evolution prescription in \cite{Bacchetta:2017gcc} and set $\zeta=\mu$. The unpolarized TMD distribution in configuration space for a parton with flavor $a$ at a certain scale $\mu^2$ is written as \cite{Bacchetta:2017gcc}
 \begin{equation}
 	\begin{aligned}
 		\tilde{f}_1^a\left(x, b_T^2 ; \mu^2\right)=& \sum_{i=q, \bar{q}, g}\left(C_{a / i} \otimes f_1^i\right)\left(x, \bar{b}_*, \mu_b^2\right) \\
 		& \times e^{S\left(\mu_b^2, \mu^2\right)}\left(\frac{\mu^2}{\mu_b^2}\right)^{-K\left(\bar{b}_* ; \mu_b\right)}\left(\frac{\mu^2}{\mu_0^2}\right)^{g_K\left(b_T\right)} \tilde{f}_{1 \mathrm{NP}}^a\left(x, b_T^2\right),
 	\end{aligned}
 \label{eq:CS}
 \end{equation}
The scale $\mu_b$ is 
\begin{equation}
	\mu_b=\frac{2 e^{-\gamma_E}}{\bar{b}_*},
\end{equation}
with $\gamma_E$ is the Euler constant and 
\begin{equation}
	\bar{b}_* \equiv \bar{b}_*\left(b_T ; b_{\min }, b_{\max }\right)=b_{\max }\left(\frac{1-e^{-b_T^4 / b_{\max }^4}}{1-e^{-b_T^4 / b_{\min }^4}}\right)^{1 / 4}.
\end{equation}
with
\begin{equation}
	b_{\max }=2 e^{-\gamma_E} / \mu_0 , \quad b_{\min }=2 e^{-\gamma_E} / \mu.
	\label{eq:bmaxbmin}
\end{equation}
The above choice guarantees that at the initial scale $\mu=\mu_0$ any
effect of TMD evolution is absent. 
At leading order of $C_{a/i}$, the evolved TMD  reduces to
\begin{equation}
	\tilde{f}_1^a\left(x, b_T^2 ; \mu^2\right)=f_1^a\left(x ; \mu_b^2\right) e^{S\left(\mu_b^2, \mu^2\right)} e^{\frac{1}{2}g_K\left(b_T\right) \ln \left(\mu^2 / \mu_0^2\right)} \tilde{f}_{1 \mathrm{NP}}^a\left(x, b_T^2\right),
	\label{eq:Sudakov_evo}
\end{equation}
where the Sudakov factor is \cite{Bacchetta:2017vzh}
\begin{equation}
	S\left(\mu_b^2, \mu^2\right)=-\frac{1}{2}\int_{\mu_b^2}^{\mu^2} \frac{d \mu^{\prime2}}{\mu^{\prime2}}\left[A\left(\alpha_S\left(\mu^{\prime2}\right)\right) \ln \left(\frac{\mu^2}{\mu^{\prime2}}\right)+B\left(\alpha_S\left(\mu^{\prime2}\right)\right)\right].
	\label{eq:Sudakov FFs}
\end{equation}
The functions $A$ and $B$ have a perturbative expansions of the form
\begin{equation}
	A\left(\alpha_S\left(\mu^2\right)\right)=\sum_{k=1}^{\infty} A_k\left(\frac{\alpha_S}{\pi}\right)^k, \quad B\left(\alpha_S\left(\mu^2\right)\right)=\sum_{k=1}^{\infty} B_k\left(\frac{\alpha_S}{\pi}\right)^k .
	\label{eq:ABfunction}
\end{equation}
To NLL accuracy, one has \cite{Davies:1984hs,Collins:1984kg}
\begin{equation}
	A_1=C_F, \quad A_2=\frac{1}{2} C_F\left[C_A\left(\frac{67}{18}-\frac{\pi^2}{6}\right)-\frac{5}{9} N_f\right], \quad B_1=-\frac{3}{2} C_F.
	\label{eq:A1A2B1}
\end{equation}
Following Refs. \cite{Nadolsky:1999kb,Landry:2002ix,Konychev:2005iy}, the non-perturbative Sudakov factor in Eq. (\ref{eq:Sudakov_evo}) is modeled as
\begin{equation}
	g_K\left(b_T\right)=-g_2 b_T^2 / 2,
	\label{eq:g2}
\end{equation}
with $g_2$ a free parameter. We choose $g_2\approx 0.13$ following the findings in \cite{Bacchetta:2017gcc,Bacchetta:2017vzh}.

In addition to the Sudakov factors, the evolved TMD functions (\ref{eq:Sudakov_evo}) are also related to two distribution functions, i.e., the collinear PDFs $f_1^a$ and  the intrinsic nonperturbative part of the TMD $\tilde{f}_{1 \mathrm{NP}}^a$. The former can be obtained by DGLAP evolution, and the latter can be determined from the condition that  Eq.~(\ref{eq:Sudakov_evo}) reduces to our calculated TMDs at initial scale, as any evolution effect should be absent at $\mu=\mu_b=\mu_0$.

The evolution of pion's Boer-Mulders function follows analogously. The Boer-Mulders function in the $b_T$-space is defined as \cite{Li:2019uhj}
\begin{equation}
	\tilde{h}_{1, q / \pi}^{\perp \alpha}(x, b_T ; \mu)=\int d^2 \bold{k_{T}} e^{-i \bold{k_{T}} \cdot \bold{b_{T}}} \frac{k_{T}^\alpha}{M_\pi} h_{1, q / \pi}^{\perp}\left(x, \bold{k}_{T}^2 ; \mu\right).
	\label{eq:BMF-Fourier}
\end{equation}
For small $b_T$ and at leading order in $\alpha_s$, it can be expressed with twist-3 correlation function \cite{Collins:1984kg,Bacchetta:2013pqa,Li:2019uhj}
\begin{equation}
	\tilde{h}_{1, q / \pi}^{\alpha \perp}(x, b_{T} ; \mu)=\left(\frac{-i b_{T}^\alpha}{2}\right) T_{q / \pi, F}^{(\sigma)}(x, x ; \mu),
	\label{eq:BMF-CS}.
\end{equation}
From Eqs. (\ref{eq:BMF 1st moment}), (\ref{eq:twist-3}), one has \cite{Li:2019uhj}
\begin{equation}
	T_{q / \pi, F}^{(\sigma)}(x, x ; \mu)=\int d^2 \bold{k_{T}} \frac{\bold{k_{T}}^2}{M_\pi} h_{1, q / \pi}^{\perp}\left(x, \bold{k_{T}}^2 ; \mu\right)=2 M_\pi h_{1, q / \pi}^{\perp(1)}.
	\label{eq:twsit3-ist}
\end{equation}
Therefore, analogous to the unpolarized TMD in Eq.~(37), the evolved Boer-Mulders function of the pion in $b_T$-space is
\begin{equation}
		\tilde{h}_{1, q / \pi}^{\alpha \perp}(x, b_T ; \mu) 
		=\left(\frac{-i b_{T}^\alpha}{2}\right) e^{S\left(\mu_b^2, \mu^2\right)} e^{\frac{1}{2}g_K\left(b_T\right) \ln \left(\mu^2 / \mu_0^2\right)} T_{q / \pi, F}^{(\sigma)}\left(x, x ; \mu_b\right)\tilde{h}_{1, q / \pi,\mathrm{NP}}^{\alpha \perp}(x, b_T ),
		\label{eq:BMF-Sudakov-evo}
\end{equation}
where $\tilde{h}_{1, q / \pi,\mathrm{NP}}^{\alpha \perp}(x, b_T )$ is the intrinsic nonperturbative part of Boer-Mulders function.  When $\mu = \mu_0$, the left hand side of Eq.~(\ref{eq:BMF-Sudakov-evo}) reduces to the calculated Boer-Mulders function at initial scale.

\begin{figure}[htbp]
	\centering  
	\subfigure{
		\label{fig:h1-evo}
		\includegraphics[width=0.47\textwidth]{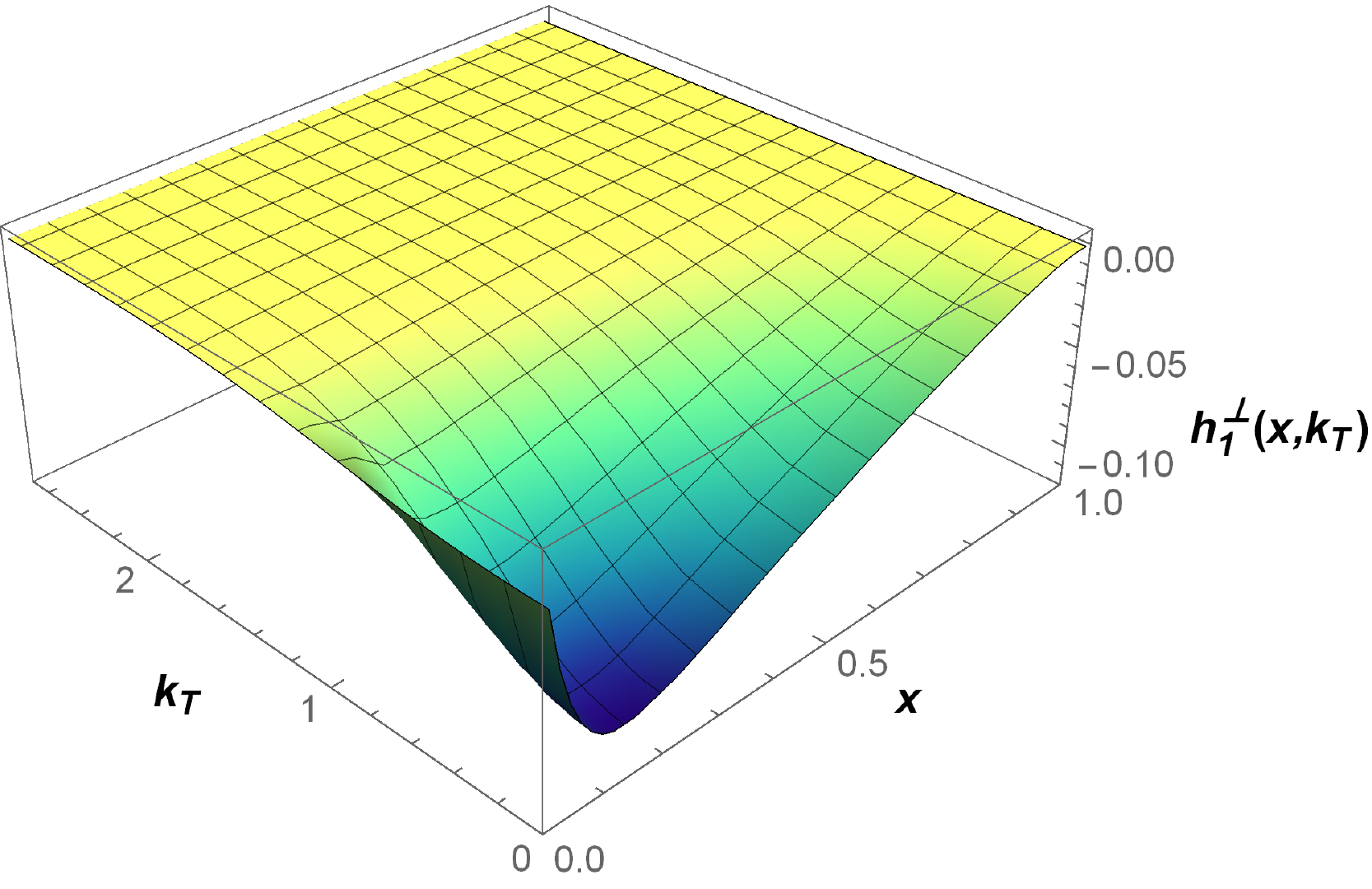}}
	\subfigure{
		\label{fig:f1-evo}
		\includegraphics[width=0.47\textwidth]{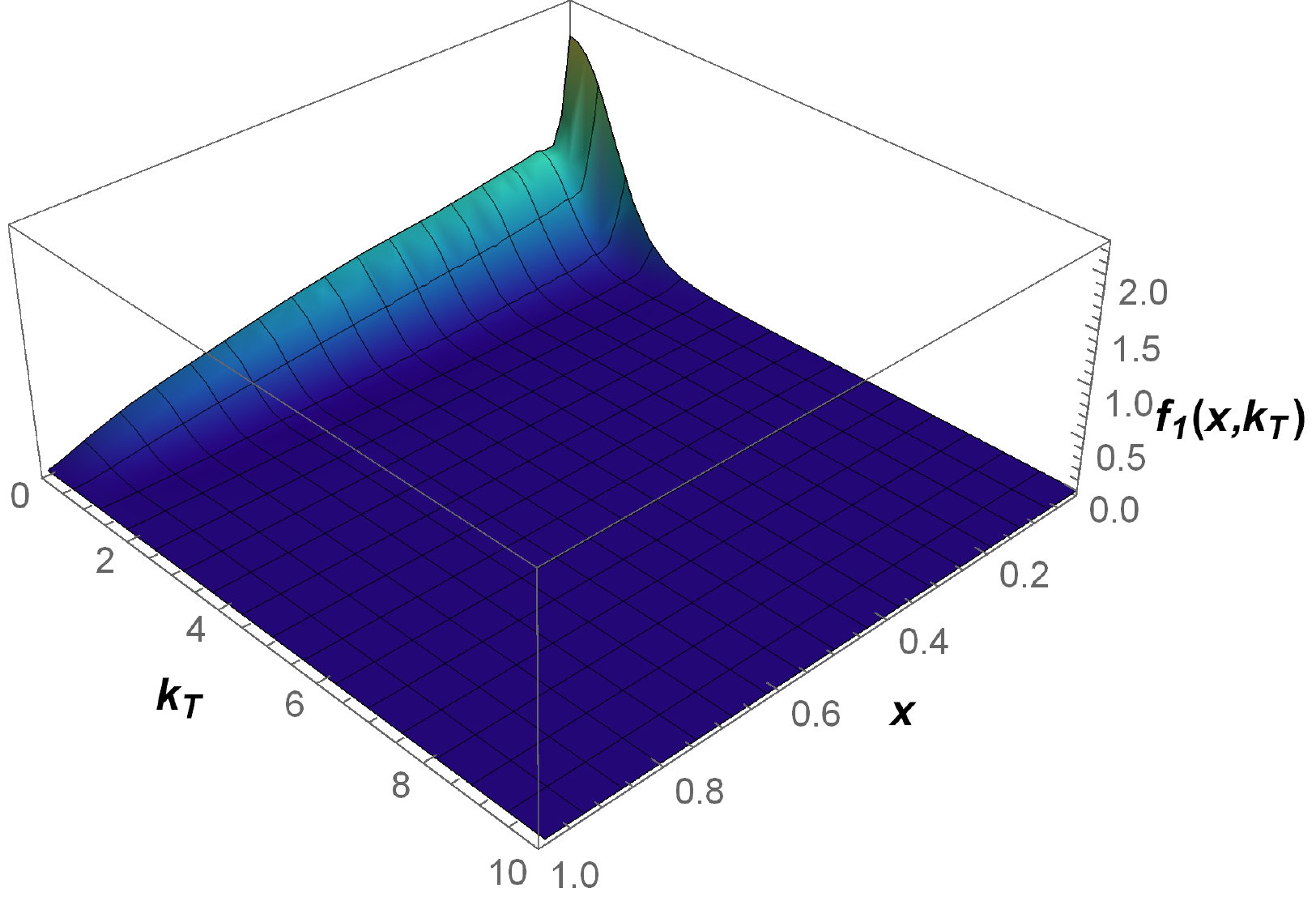}}
	\caption{(left) The evolved Boer-Mulders function generated by the non-perturbative kernel at $\mu=2.0$ GeV. (right) The evolved unpolarized TMD $f_1$ at $\mu=2.0$ GeV.}
	\label{fig:TMDs-evo}
\end{figure}

After evolving the TMDs in $b_T$-space, we Fourier transform the TMDs back to $k_T$ space. In FIG. \ref{fig:TMDs-evo}, we show the unpolarized TMD and Boer-Mulders function of pion evolved to the scale of $\mu=2.0$ GeV. As compared to TMDs at hadronic scale in FIG.\ref{fig:TMDs}, these TMDs shift toward lower $x$ and gets broader in $k_T$. 

\begin{figure}[H]
	\centering
	\includegraphics[width=0.6\textwidth]{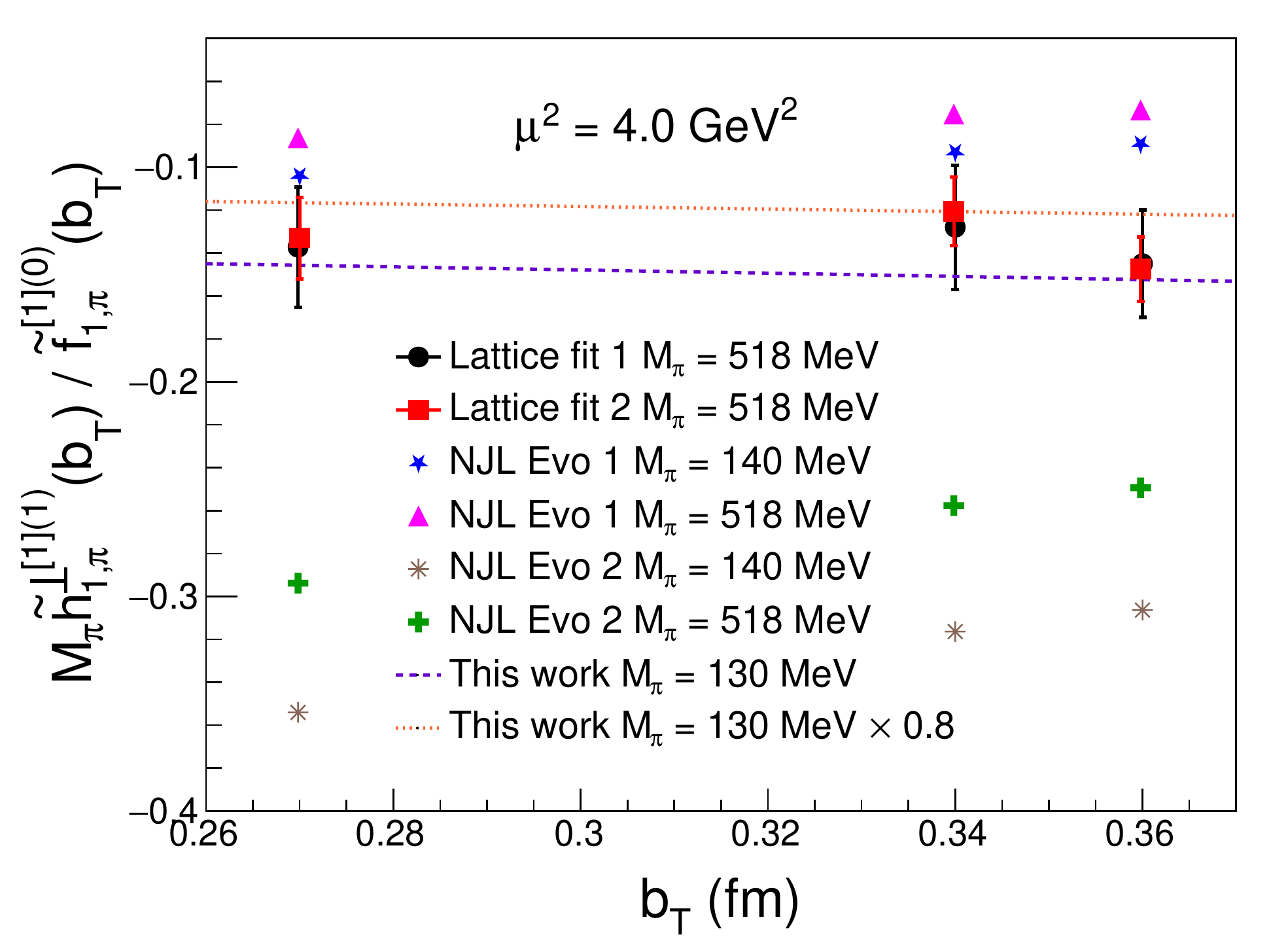}
	\caption{The generalized Boer-Mulders shift, Eq. (\ref{eq:BM-shift}), as a function of $b_T$. The results of NJL model include two types of evolution schemes, Evo 1 and 2, corresponding to different evolution factors (see Ref. \cite{Noguera:2015iia}). We indicate the generalized Boer-Mulders shift calculated using LFWFs, corresponding to pion mass $M_\pi=130$ with dashed line in the figure. The dotted line indicates the result of the dashed line multiplied by 0.8. Two sets of lattice data (Lattice fit 1 and 2), with
		their RMS deviation, obtained in Ref. \cite{Engelhardt:2015xja} through independent fits, are shown for comparison.
	}
	\label{fig:BMF-shift}
\end{figure}

Finally, we compare our results with lattice calculation, focusing on a quantity named the generalized Boer-Mulders shifts \cite{Engelhardt:2015xja}. For pion it has been predicted using a pion of mass $M_\pi=518$ MeV at the scale of 2 GeV \cite{Engelhardt:2015xja}, which is  defined as 
\begin{equation}
	\left\langle k_{T}\right\rangle_{U T}\left(b_{T}^2; \mu^2=4.0\ \mathrm{GeV}^2\right)=M_\pi \frac{\tilde{h}_{1, \pi}^{\perp[1](1)}\left(b_{T}^2;\mu^2=4.0\ \mathrm{GeV}^2\right)}{\tilde{f}_{1, \pi}^{[1](0)}\left(b_{T}^2;\mu^2=4.0\ \mathrm{GeV}^2\right)},
	\label{eq:BM-shift}
\end{equation}
where the generalized TMD moments are given by 
\begin{equation}
		\tilde{f}^{[m](n)}\left(b_{T}^2\right)= \frac{2 \pi n !}{M_\pi^{2 n}} \int \mathrm{d} x x^{m-1} \int \mathrm{d} k_{T} k_{T}\left(\frac{k_{T}}{b_{T}}\right)^n 
		 J_n\left(b_{T} k_{T}\right) f\left(x, k_{T}^2\right),
		 	\label{eq:GTMDs moment}
\end{equation}
and analogously for $\tilde{h}_{1, \pi}^{\perp[1](1)}$. The lattice prediction is displayed as the data points with error bars. Our result is shown in FIG. \ref{fig:BMF-shift} as the purple dashed line.  In addition, we also include the results from the NJL model \cite{Noguera:2015iia}. We find that in NJL model, the results of $M_\pi=518$ MeV in approximately 0.8 times that of 140 MeV, as indicated by both evolutions \cite{Noguera:2015iia}. Thus we multiply the purple curve by 0.8 and obtain the red dotted line as an estimate of pion mass variation effect. We find our results agree quite well with lattice calculation \cite{Engelhardt:2015xja}. 
Considering the uncertainty in the non-perturbative Sudakov factor, we test with different $g_2$ values in Eq.~(\ref{eq:g2}) and find the calculated generalized Boer-Mulders shift to be insensitive to $g_2$. This is reasonable as the ratio in Eq.~(\ref{eq:BM-shift}) is devised to cancel the soft factors in the numerator and denominator.

\section{Conclusions}
\label{sec:conclusion}
In this work, we study the unpolarized TMD and Boer-Mulders function of pion using DS-BSEs based LFWFs. The light front overlap representation is employed. To obtain a nonvanishing Boer-Mulders function, final-state interaction analysis is utilized to construct the gluons rescattering kernel $iG(x,q_T)$, which incorporates the information of gauge link in SU(3) case. The two leading twist TMDs of pion at hadronic scale $\mu_0 \approx 0.5$ GeV are thus given.  We then evolve both TMDs to a higher scale of $2$ GeV. The generalized Boer-Mulders shift is found to be in good agreement with lattice prediction.  We remark that at leading Fock-state, the Boer-Mulders function is proportional to the pion's p-wave LF-LFWFs. Its emergence and magnitude is thus a reflection of the p-wave components inside pion, which essentially arises from the relativistic internal motion inside the pion.  The DS-BSEs approach encapsulate such property, as well as QCD's dynamical chiral symmetry breaking property, and pass them onto the DS-BSEs based LF-LFWFs, and eventually the pion TMDs.

\begin{acknowledgments}
We thank Xiaoyu Wang for beneficial communications about QCDNUM program. This work is supported by the National Natural Science Foundation of China (Grant No. 11905104) and the Strategic Priority Research Program of Chinese Academy of Sciences (Grant NO. XDB34030301).
\end{acknowledgments}

\bibliographystyle{apsrev4-1}
\bibliography{pion_boer_mulders_function}
\newpage

\end{document}